# Chip-scale sensor for spectroscopic metrology

Chunhui Yao[1]†, Wanlu Zhang[1]†, Peng Bao[1], Jie Ma[2], Wei Zhuo[2], Minjia Chen[1], Zhitian Shi[1], Jingwen Zhou[2], Yuxiao Ye[2], Liang Ming[2], Ting Yan[2], Richard Penty[1], Qixiang Cheng[1,2]*

1. Electrical Engineering Division, Department of Engineering, University of Cambridge, UK
2. GlitterinTech Limited, Xuzhou, China
*qc223@cam.ac.uk;
†These two authors contribute equally to this work.

**Abstract**

Miniaturized spectrometers hold great promise for in situ, in vitro, and even in vivo sensing applications. However, their size reduction imposes vital performance constraints in meeting the rigorous demands of spectroscopy, including fine resolution, high accuracy, and ultra-wide observation window. The prevailing view in the community holds that miniaturized spectrometers are most suitable for the coarse identification of signature peaks. In this paper, we present an integrated reconstructive spectrometer that enables near-infrared (NIR) spectroscopic metrology, and demonstrate a fully packaged sensor with auxiliary electronics. Such a sensor operates over a 520 nm bandwidth together with a resolution of less than 8 pm, which translates into a record-breaking bandwidth-to-resolution ratio of over 65,000. The classification of different types of solid substances and the concentration measurement of aqueous and organic solutions are performed, all achieving approximately 100% accuracy. Notably, the detection limit of our sensor matches that of the commercial benchtop counterparts, which is as low as 0.1% (i.e. 100 mg/dL) for identifying the concentration of glucose solution.

**Main**

Near-infrared (NIR; 780 nm to 2500 nm) spectrometry, one of the most essential vibrational spectroscopy techniques, is widely applied in numerous fields such as biomedicine, chemistry, and material science. Over the past decade, the surging demand for in situ, in vitro, and in vivo NIR spectroscopic analysis, including wearable devices for healthcare monitoring, portable tools for chemical detection, and compact optical systems for hyperspectral imaging, has driven the development of miniaturized spectrometers towards smaller sizes and higher performance. Leveraging the compatibility with mature CMOS technologies, silicon photonics offers a low-cost platform for developing chip-size spectrometers[1]. Additionally, the broad transparency windows of Si (from 1.1 μm to 8 μm) and SiN (from 370 nm to 5 μm) perfectly cover the spectral range for NIR detection[2]. Currently, most on-chip spectrometer designs adhere to the same principles as traditional dispersive or Fourier transform (FT) spectrometers but leverage photonic integration technologies to obtain a millimeter scale footprint[3]. The size reduction, however, in return sets limitations to the dispersive or interferometric elements, which fundamentally bound their bandwidth and resolution [4,5]. In recent years, the emergence of reconstructive spectrometers (RSs) with compressive sensing algorithms has paved a new way for miniaturized spectrometers [6]. In principle, RSs employ a global sampling strategy to resolve the entire incident spectrum, generally by forming an underdetermined matrix for extensive information acquisition. This feature makes them appealing to be implemented on chip, given their simplicity. However, the notable dispersion effects in the integration platform impose substantial constraints, especially on the operational bandwidth, either for passive devices[7–9], or active designs[10–13].

In contrast, NIR spectroscopy applications necessitate fine resolution, high accuracy, and wide working bandwidth for capturing informative spectral features[14]. Figure 1a summarizes the overtone absorption bands of typical functional groups in the NIR range[15]. The results clearly show that: 1) a functional group typically exhibits multiple overtone bands that span over hundreds of nanometers, and 2) the overtone bands of different functional groups overlap. The above attributes could largely complicate spectroscopic analysis, due to potential data underfitting [16]. For instance, the metrological measurement of typical biomarkers such as glucose, lactate, and urea, demands tracking various organic functional groups, including -OH, -CH and -$CH_2$, thus requiring an ultra-wide observation window of hundreds of nanometers[17,18]. Likewise, the spectroscopic analysis of solid, liquid or gas substances for pharmaceutical, agriculture, fuel, and atmospheric monitoring may even need a functional bandwidth of over a thousand nanometers[19]. Figure 1b summarizes the bandwidth and resolution specifications adopted in various NIR applications for biomedical sensing[20–26] and industrial chemical detection[27–32], along with the performance metrics of state-of-the-art chip-scale RSs[7,8,10–13,33–40], where there remains a clear gap. It is thus commonly accepted that miniaturized spectrometers are more for the identification of signature spectral peaks rather than quantitative metrology[41].

In this paper, we present an ultra-high-performance integrated RS that empowers NIR spectroscopic metrology. The RS chip simply consists of a cascade of tunable micro-ring resonators (MRRs), which exploit a dispersion-engineered directional coupler (DC). Its sampling response is temporally decorrelated by manipulating the phase of each MRR. This chip is implemented on a SiN integration platform, and gets fully packaged into a chip-scale sensor with auxiliary electronics, operating over a 520 nm bandwidth from 1180 nm to 1700 nm. With a superior resolution of below 8 pm, this device achieves a bandwidth-to-resolution ratio exceeding 65,000, which, to the best of our knowledge, is at-least severalfold higher than any reported miniaturized spectrometer. A series of NIR spectroscopic applications is demonstrated, including the classification of plastic and coffee samples and the concentration measurement of aqueous and organic solutions, all achieving approximately 100% accuracy. Most importantly, the detection limit of our sensor is examined using glucose solution with concentrations as low as 0.1% (i.e. 100 mg/dL or 5.55 mmol/L) identified. Such level of detectability is already comparable to the results obtained from commercial benchtop spectrometers, establishing a new benchmark for NIR spectroscopy with miniaturized sensors.

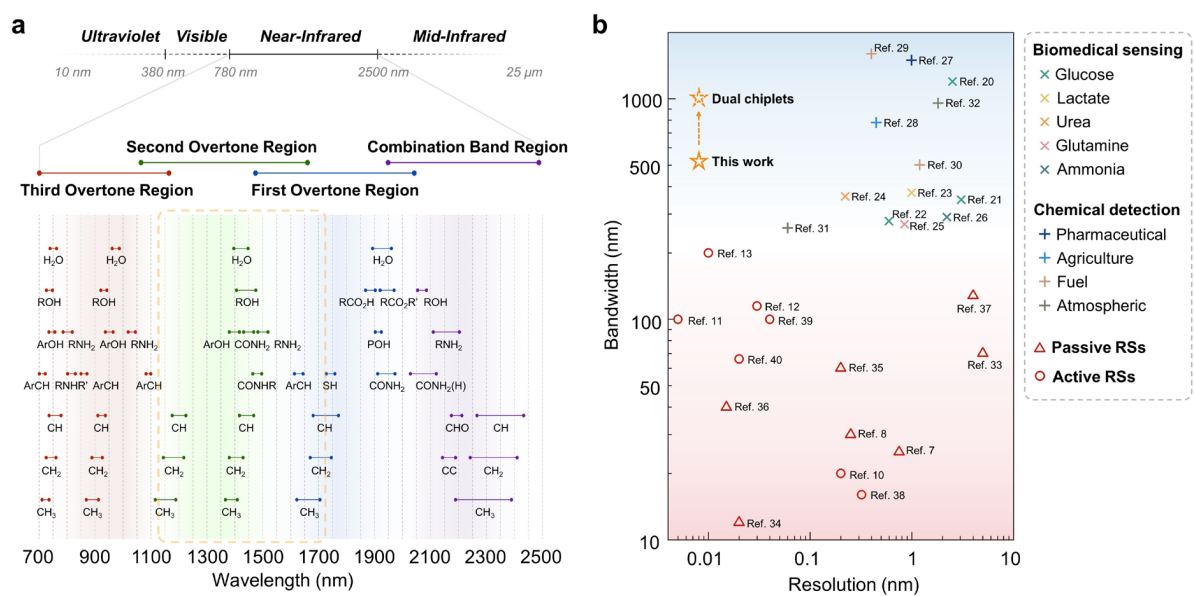

**Figure 1 | NIR spectroscopy needs vs. RS performance.** (a) Overtone absorption bands of typical functional groups in the NIR spectrum range. (b) Performance requirements of various NIR spectroscopy applications compared with the metrics of state-of-the-art waveguide-based RSs. The colored dots represent specific application scenarios. The dashed arrow indicates that sensor performance can be further enhanced by co-packaging two or even more chiplets (see Discussion and Conclusion for details).

## Results

### Principle and design

The working principle of RSs is elaborated in Methods. As revealed by the compressive sensing theory, an RS necessitates a sufficient number of sampling channels with rapid and random spectral perturbations, also known as spectral speckles[42]. To achieve this, we propose an ultra-broadband single-bus RS based on a cascade of dispersion engineered MRRs, as illustrated by Fig. 2a. By opting for the SiN platform, our device benefits from a material dispersion that is over four times lower than that of the Si-on-insulator (SOI)[2]. Figure 2b presents the MRR's structural diagram in a vertical racetrack style. The transmission spectrum of such an MRR can be generally written as[43]:

$$T_i = \frac{\alpha^2 + r^2 - 2\alpha r \cos(\varphi_i + \delta_i)}{1 + \alpha^2 r^2 - 2\alpha r \cos(\varphi_i + \delta_i)} \quad (1)$$

where $r$ is the self-coupling coefficient, $\alpha$ denotes the power attenuation coefficient, $\varphi_i$ is the single-pass phase shift of the ring, and $\delta_i$ represents the external phase tuning. Based on Eq. 1, the transmission intensity at resonance wavelength $T_{res}$ can be derived as:

$$T_{res} = \frac{(r-\alpha)^2}{(1-r\alpha)^2} \quad (2)$$

Instead of operating under the stringent critical coupling condition as high-Q narrowband filters, the MRRs in our design are tailored to be over-coupled and exhibit moderate extinction ratios, in order to optimize overlaid spectral fluctuations with minimal excess loss. The discrepancies in the FSR of individual MRRs caused by dispersion are specifically harnessed to increase spectral randomness and reduce the correlation between sampling channels. These features afford our design a high degree of flexibility. Figure 2c shows the calculated $T_{res}$ under different combinations of $\alpha$ and $r$. The inset highlight the region where an MRR can maintain the over-coupling state while keeping the $T_{res}$ in a preferable range. It is thus crucial to strike a balance between the loss coefficient α and the coupling efficiency throughout a maximal operational bandwidth.

For this purpose, we strategically deploy a pair of waveguide bends with a fixed but relatively small radius ($R_2$) to make its bending loss dominate in the ring, so that its round-trip loss naturally increases with longer wavelengths. Meanwhile, we adopt a curved DC structure for greater wavelength stability, thanks to its enhanced phase-matching capability[44]. Its structural parameters are globally optimized using the particle swarm optimization (PSO) algorithm to obtain an increasing cross-coupling efficiency over wavelength, thus matching the increased on-ring loss. Detailed design process is described in Methods. Figure 2d,e present the finite-difference time-domain (FDTD) simulated coupling efficiency and light propagation profiles of the tailored curved DC, respectively, showing a rising cross-coupling ratio from 1200 nm to 1700 nm. Besides, each MRR is equipped with a pair of straight waveguides with varying lengths ($L_i$) to not only achieve small FSRs for rapid spectral roll-offs, but also to break any periodicities in the overlaid response. Additionally, the MRRs incorporate thermo-optic (TO) phase shifters to temporally decorrelate the sampling responses. By tuning the MRR's phase (i.e. $\delta_i$) into multiple states between 0 to $2\pi$, the cascading system yields an exponentially scalable number of sampling channels, which equals to the cumulative product of phase states at each MRR. This design, therefore, effectively produces spectral sampling matrices with thousands of temporal channels.

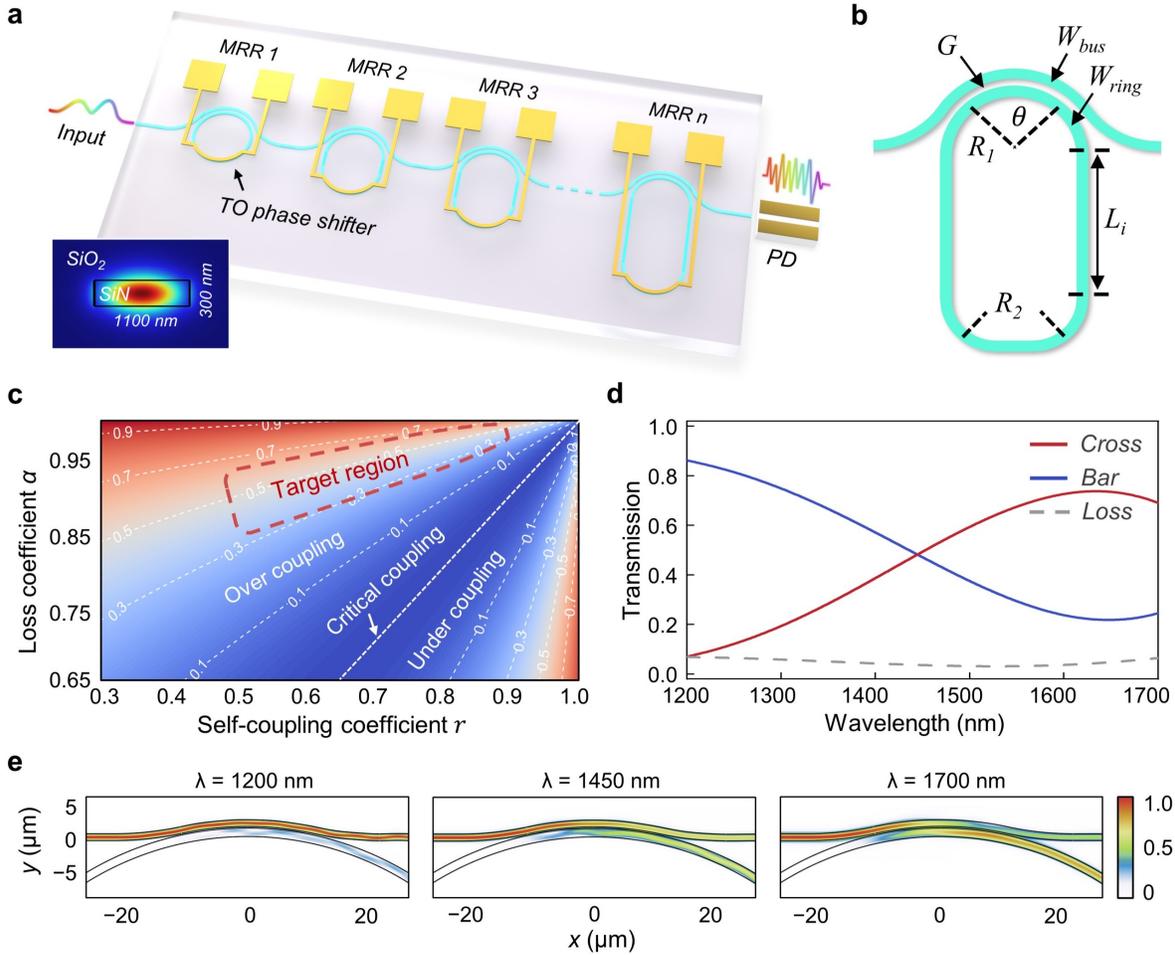

**Figure 2 | Spectrometer design and simulations.** (a) Conceptual Schematic of the proposed ultra-broadband spectrometer featuring multiple stages of MRRs on a single bus. The inset shows the mode field distribution on the bus waveguide. (b) Structural diagram of a dispersion engineered MRR with a curved DC. (c) Calculated values of $T_{res}$ under different combinations of $\alpha$ and $r$. The insets highlight the target region of $\alpha$ and $r$ for the MRR filters. (d) Simulated coupling efficiency of the optimized curved DC, demonstrating an increasing coupling efficiency over wavelength. (e) Simulated light propagation profiles (i.e. electric field intensity) in the curved DC at different wavelengths.

**Experimental characterization**

Figure 3a presents our fully packaged NIR spectroscopic sensor at centimeters scale, incorporating the SiN RS chip and a high-speed driving board with a microcontroller unit (MCU) integrated. The insets enlarge two optical sampling interfaces that are tailored for reflective and transmissive measurements, respectively. More details regarding the driving board and sampling interfaces are provided in Methods. The RS chip comprises six cascaded MRRs as a balance between performance and complexity. Figure 3b details the device packaging, where the chip is wire-bonded for electrical fanout and optically accessed via lensed polarization-maintaining fibers (PMFs). A microscope image of the RS chip is shown in Fig. 3c, with the insets magnifying three tunable MRRs with curved DC. The MRRs each occupies a footprint of less than 80 × 150 µm² and are laterally spaced by 200 µm to minimize thermal crosstalk. We set four phase states per MRR, creating a total of 4096 temporal sampling channels. Figure 3d plots representative measured sampling channels over a 520 nm bandwidth between 1180 nm to 1700 nm (Supplementary Fig. S1 shows the whole sampling matrix). The insets further display the channel responses at four different observation windows.

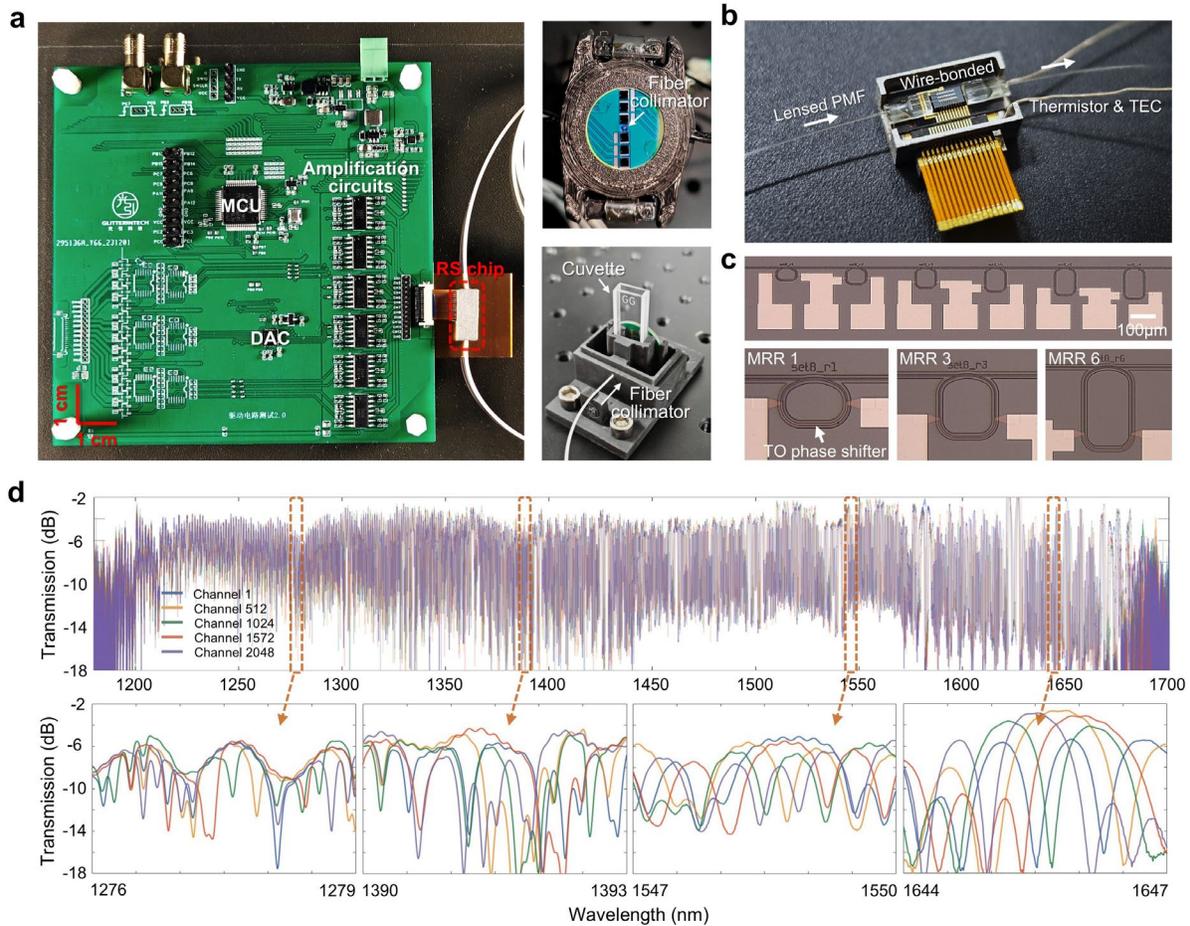

**Figure 3 | Device images and measured channel spectral responses.** (a) Photograph of the miniaturized NIR spectrometric sensor. The insets show the optical sampling interfaces for measuring reflection and transmission spectra, respectively. (b) Enlarged image of the fully packaged spectrometer chip. (c) Microscope image of the fabricated RS with six stages of tunable MRRs. The insets provide enlarged views of MRRs with different circumferences. (d) Representative examples of the measured channel spectral responses between 1180 nm and 1700 nm. The insets depict different observation windows, highlighting the rapid pseudo-random spectral fluctuations.

To characterize the RS performance, we first test a variety of discrete incident spectra, including single-, dual-, and triple-peak laser signals at different wavelengths. The reconstruction algorithm and processes are detailed in Methods. The reconstruction accuracies are quantified by the *L2-norm* relative errors $\varepsilon$, defined as $\varepsilon = \|\Phi_0 - \Phi\|_2 / \|\Phi_0\|_2$ where $\Phi_0$ and $\Phi$ denote the reference and reconstructed spectra, respectively. As shown in Fig. 4a-c, our RS precisely resolves the intensities and locations for all peaks, exhibiting low relative errors ranging from 0.04 to 0.11. Notably, the spectral spacing of dual-peak signals is gradually reduced down to 8 pm, marking its resolution according to the Rayleigh criterion. Rigorous simulations further show that such a sensing resolution can be well achieved across the entire operational bandwidth (see Supplementary Fig. S2). Subsequently, various continuous, broadband spectra that are generated by a benchtop waveshaper are tested. Figure 4b depicts that all complex waveform features are well reconstructed. Hybrid incident spectra have also been measured, with different locations of a laser peak combining the ASE spectra of an EDFA. Figure 4e shows that both the broadband and narrowband spectral components are well distinguished for all cases.

The global sampling feature of our RS offers the flexibility to customize the number of sampling channels in any individual measurement, allowing a trade-off between the reconstruction accuracy, measurement time, and computational complexity to enable user-definable performance metrics[13]. Accordingly, we investigate the impact of the channel number on reconstruction performance, covering all cases of single-peak, dual-peak, and broadband inputs. Here, the spectral resolution is scaled down in accordance with the decrease of sampling channels to maintain a consistent compression ratio. Figure 4f plots the reconstruction error and computing time as a function of the channel number, and highlights the reconstructed broadband spectra under different channel numbers. As can be seen, while reducing the channel number gradually induces higher relative error, it sharply lowers the computational cost. On the other hand, the impact of spectral window width is explored, as shown by Fig. 4g, revealing that reducing the redundant bandwidth in reconstruction leads to clear accuracy enhancement. This is attributed to the noise accumulated over the signal bandwidth that does not carry inputs (see Methods for detailed strategies on adjusting the reconstruction window). Furthermore, we examine the temperature stability of our RS. Measurements reveal that the sampling response redshifts by only 0.32 nm when the ambient temperature increases from 10 °C to 30 °C, which can be easily handled by tweaking the reconstruction matrix and potentially enables a cooling-free operation. This is also translated into a temperature tolerance of around ± 0.75 °C (see Supplementary Fig. S3).

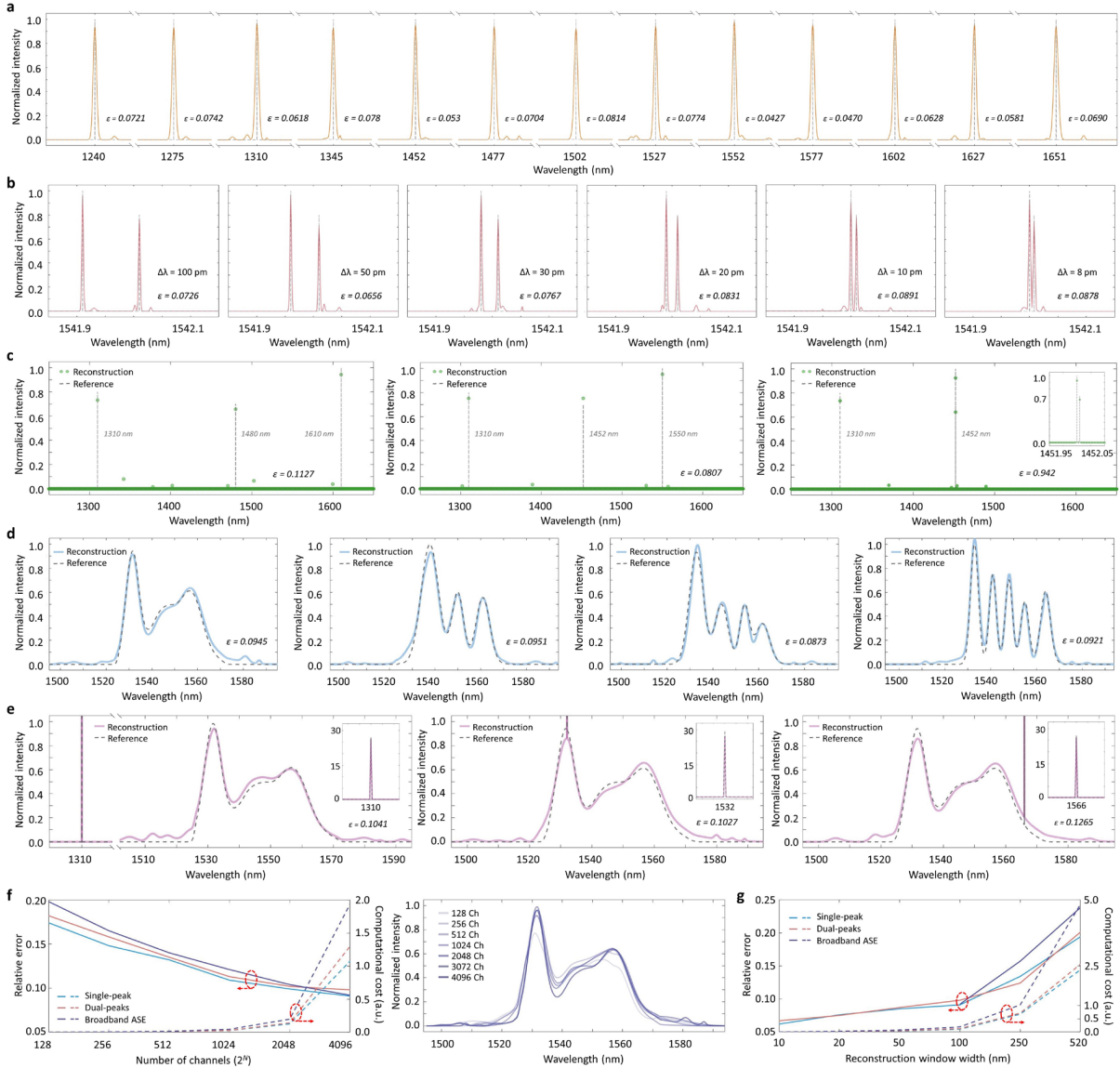

**Figure 4 | Performance characterization of the RS.** (a) Reconstructed spectra for a series of single-peak laser signals. The black dashed lines indicating their center wavelengths. (b) Reconstructed spectra for dual-peak laser signals with decreasing spectral spacing from 100 pm down to 8 pm. (c) Reconstructed spectra for triple-peak laser signals at different wavelength positions, with the spectral spacings between peaks ranging from as close as 10 pm to over 200 nm. (d) Reconstructed broadband signals generated by the ASE of an EDFA, filtered by a commercial waveshaper with various spectral patterns. (e) Reconstructed spectra of hybrid signals containing both broadband and narrowband spectral components. The inset shows the reconstruction of the narrowband laser peaks at different wavelengths. (f) Left: investigation of reconstruction accuracy and computing time versus sampling channel number. The solid lines represent the reconstruction accuracies, while the dashed lines denote the computational costs. Right: the reconstructed broadband spectra using different numbers of sampling channels. (g) Relationship between the reconstruction accuracy and computing time versus the width of reconstruction window for various types of input signals (with channel number fixed at 512). Note that the minimal window width for broadband ASE signal is restricted to 100 nm to prevent any information loss.

**Applications for NIR spectrometric sensing**

With the fully-packaged NIR sensor, we showcase a series of bandwidth-demanding spectroscopic applications for material classification and solution concentration monitoring, addressing typical industrial requirements[1]. We examine ten different types of plastics commonly used in chemical manufacturing and ten varieties of coffee powers originated from diverse regions. These samples are visually indistinguishable, but exhibit subtle spectral differences over the NIR range. We repeatedly measure each sample for 60 times over the entire 520 nm bandwidth, and randomly split the data into training and testing sets in a 7:3 ratio. All measurements only take 312 sampling channels considering that the spectra under test have modest roll-off ratios. Figure 5a,b show the measured reflectance of two representative samples, respectively, for the plastic and coffee. The reflectance spectra for all remaining samples are provided in Supplementary Fig. 4. In these measurements, four superluminescent diodes (SLDs) with different emitting wavelengths are utilized to maximize the bandwidth of illumination. The minor discontinuities in the measured spectra are due to the low SLD power density at those wavelength regions (see more discussions in Methods). Corresponding classification models are trained on the basis of support vector machine (SVM)[45] to identify plastic or coffee samples. Figure 5c,d present the classification results for the plastic and coffee samples, respectively, all demonstrating 100% accuracy.

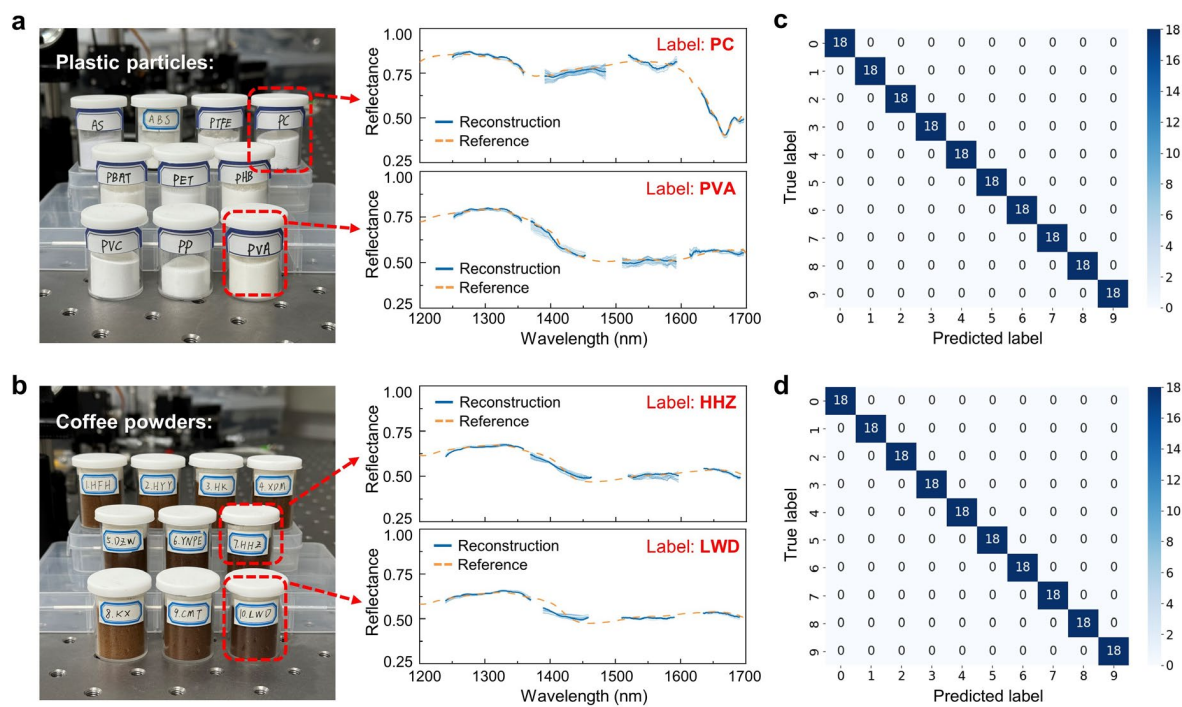

**Figure 5 | Classification of various solid substances.** (a,b) Reflectance measurement of ten different types of plastic and coffee samples, respectively. The insets show the measured reflectance spectra of two representative plastic or coffee samples (each repeated for 60 times), respectively, revealing superior measurement repeatability of our NIR sensor. The orange dashed lines represent the references measured using a commercial benchtop spectrometer. One of the repetitions is highlighted to indicate the high reconstruction accuracy. (c,d) Confusion matrix of the classification results for various plastic or coffee samples, respectively.

The concentration monitoring for various aqueous and organic solutions is further conducted. We test the ethanol aqueous solutions, solutions of ethylene glycol (EG) in isopropanol (ISO), and glucose aqueous solutions. Figure 6a shows the measured absorptance spectra of ethanol solutions, which, despite the strong water absorption bands, display the unique NIR features of ethanol. For example, from 1640 nm to 1700 nm, the measured absorptance first decreases and then increases as the ethanol concentration rises, with a turning point at around 1675 nm. This distinctive signature matches the results from reported studies[46], and can be explained by the absorption of the C-H bonds in ethanol (see Fig. 1a). Accordingly, we train a prediction model using the random forest (RF) algorithm (see Methods for modelling details)[47], which achieves a coefficient of determination ($R^2$) of 1.000 on the testing sets. Figure 6b presents a scatter plot of the measured and predicted ethanol concentrations. The calculated mean absolute error (MAE) and root mean squared error (RMSE) for the testing sets are only 0.057% and 0.074%, respectively. Figure 6c presents the measured absorptance spectra for EG solutions in ISO, showing clear concentration-induced variations in absorption. It is worth noting that distinguishing EG from ISO is crucial, since EG can cause severe renal failure and is thus more hazardous than ISO[48,49]. For instance, across the broad wavelength range between around 1400 nm to 1680 nm, a significant rise in absorptance can be observed as the concentration of EG increases, which is attributed to that EG has one more -OH group than ISO. Figure 6d shows the prediction results using our RF model, with the MAE and RMSE being only 0.029% and 0.058%, respectively. The measured absorptance spectra of various glucose solutions are shown in Fig. 6e. As can be seen, the spectral variations are more pronounced in the wavelength ranges of below 1400 nm and above 1600 nm, where the water absorption is minor and the overtone vibrations of glucose molecule's C-H bonds dominate[50]. Figure 6f compares the actual glucose concentrations against the predictions from our model. The calculated MAE and RMSE are 0.053% and 0.073%, respectively, illustrating the system detection limit of at-least 0.1% (i.e. 100 mg/dL or 5.55 mmol/L).

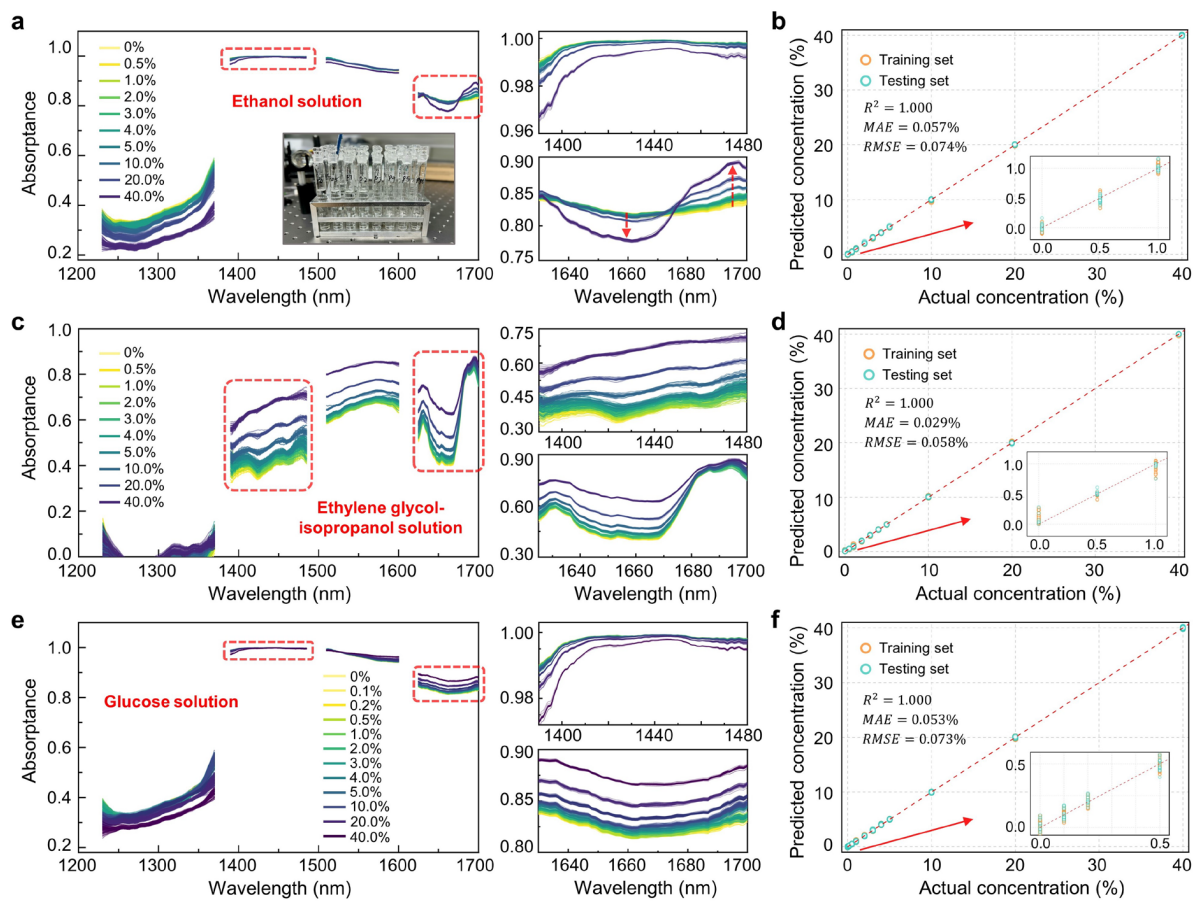

**Figure 6 | Concentration testing of various solutions.** (a) Measured absorptance spectra for ethanol aqueous solutions with concentrations decreasing from 40% to 0.5% (each repeated for 60 times). The inset shows the test tubes containing the respective solutions. The red arrows denote the trend of absorptance variation with increasing solution concentration. (b) Predicted ethanol concentrations using our RF model. The inset provides an enlarged view of the prediction results at low concentrations. (c) Measured absorptance spectra for solutions of ethylene glycol in isopropanol with concentrations decreasing from 40% to 0.5%. (d) Predicted concentrations using our RF model. (e) Measured absorptance spectra for glucose aqueous solutions with concentrations decreasing from 40% down to 0.1%. (f) Predicted glucose concentrations using our RF model.

**Discussion and conclusion**

We replicate the glucose concentration tests using benchtop dispersive and FT spectrometers (IdeaOptics NIR17+Px, and Bruker MPA II). Note that here we construct different prediction models using their full or partial operational bandwidth for fair comparisons (see Supplementary Fig. 5. for more details). Table 1 details that our NIR sensor reaches the detection limit and accuracy comparable to those of commercial counterparts.

Since only MRRs are used to shape the channel sampling response, our design allows the flexible migration of operational wavelength band. For instance, MRRs with dispersion-engineered DC targeting the waveguide range of between 700 nm and 1200 nm is readily optimized, as shown by Supplementary Fig. 6. Hence, simply by co-packaging two or more RS chiplets, a miniaturized NIR sensor with over one thousand nanometer bandwidth can be easily achieved.

In summary, we present an RS-based NIR spectroscopic sensor, achieving an > 520 nm operational bandwidth with a < 8 pm resolution. Various spectroscopic applications are examined, achieving accuracies of approximately 100%. We also demonstrate the detection limit of our sensor to be at-least 0.1% (i.e. 100 mg/dL), which is comparable to the results obtained from commercial benchtop spectrometers, representing as a big leap towards spectroscopic metrology with miniaturized spectrometers.

**Table 1. Performance Comparison based on the concentration test of glucose solutions**

| Principle | Spectral window | Detection limit | MAE | RMSE |
|---|---|---|---|---|
| Dispersion (IdeaOptics NIR17+Px) | 900 nm to 1700 nm | 0.1% | 0.022% | 0.029% |
| Fourier transform (Bruker MPA II) | 866 nm to 2500 nm | 0.05% | 0.002% | 0.011% |
| Dispersion (IdeaOptics NIR17+Px) | 1180 nm to 1700 nm | 0.1% | 0.026% | 0.035% |
| Fourier transform (Bruker MPA II) | 1180 nm to 1700 nm | 0.1% | 0.013% | 0.024% |
| Reconstruction (this work) | 1180 nm to 1700 nm | 0.1% | 0.053% | 0.073% |

## Methods

### Parameter optimization of curved DC

We utilize ANSYS Lumerical FDTD to perform the physical simulations of the curved DC and employ a custom PSO algorithm script to global optimize the DC geometry[69]. In specific, we first conduct a coarse sweep of all structural parameters to identify a relatively optimal combination, which serves as the initial parameter set. The power intensities at both output ports of the DC are used to calculate the objective function, aiming to achieve a gradually increasing cross-coupling efficiency over the widest possible wavelength range while maintaining a minimal insertion loss. Based on this, the PSO algorithm continuously iterates to obtain the optimal parameter set. The detailed structural parameters of our curved DC are listed in Supplementary Table 1.

### Chip fabrication and packaging

The spectrometer chip is fabricated via a CORNERSTONE SiN multi-project wafer run, employing standard deep ultraviolet (DUV) lithography with a feature size of 250 nm. It consists of a 300 nm thick SiN layer sandwiched between a 3 µm buried oxide layer and a 2 µm Silicon dioxide top cladding layer. The chip is diced, polished, and wire-bonded to customized PCB broad for electrical fan-out (see Fig. 3b). Lensed PMFs are used to match the waveguide edge couplers on both edges for efficient optical assessment. UV-curable adhesive is applied to mechanically secure the PMFs for stable optical alignment, measuring a coupling loss of around 2.5 dB per facet.

### Electrical control and thermal stabilization

An automatic electrical driving board is developed to enable the swift modulation of channel sampling responses, as shown by Fig. 3a. The MCU is programmed to produce the initial control signals and send them to a high-resolution multi-channel digital-to-analog converter (DAC). The output analog signals from the DAC are then amplified by amplification circuits and injected into the spectrometer chip for temporal channel sweeping. This system allows for a sampling speed of over 500 Hz, such that the measurement for each spectrum can be completed within a few seconds.

A thermistor is attached to the top surface of the chip, while a thermoelectric cooler (TEC) is placed underneath it (see Fig. 3b), jointly forming a negative feedback mechanism to facilitate the temperature stabilization during experimental testing.

### Optical testbed and sampling interfaces

To calibrate our ultra-broadband RS chip, we sequentially introduce four SLDs with different center wavelengths as inputs and measure the corresponding output spectra using a commercial benchtop spectrum analyzer. The results are then compiled to form the full-band spectral response. Likewise, during the NIR spectrometric sensing for various substances, these SLDs are also sequentially turned on to illuminate the samples. The reflected or transmitted power is then collected by photodiodes (PDs) and used for spectral reconstruction. After normalization with standard references (e.g., a fully reflective whiteboard or an unloaded cuvette), the reflectance or absorptance of the samples across the full bandwidth can be obtained. Supplementary Fig. 7 details the schematic of the testbed workflow and the ASE spectra from the SLDs, respectively.

To efficiently collect the reflected or transmitted light from samples, we develop two free-space optical sampling interfaces, as shown in the insets in Fig. 3a. Supplementary Fig. 8 details their structure and working principles, respectively. The first interface measures the reflectance of solid substances using a single-mode fiber collimator with multiple 2 mm photodiodes (PDs). The sample is positioned above the collimator, ensuring the emitted light reaches and reflects off the sample before being collected by the PDs This process allows a coupling loss of less than 8 dB. The second interface is designed for measuring the absorptance of liquid samples, where the single-mode fiber collimator is aligned with a cuvette that contains the sample. The transmitted light through the solution is captured using surface PDs. When the cuvette is empty, this interface exhibits a coupling loss of less than 1 dB.

**Reconstruction algorithm and process**

Mathematically, the underlying principle of RSs can be described using an undetermined equation, written as[4,56]:

$$I_{M\times 1} = S_{M\times N}\Phi_{N\times 1} \tag{3}$$

where $\Phi_{N\times 1}$ denotes the discretized vector of an unknown incident spectrum with $N$ spectral pixels, $S_{M\times N}$ represents the sampling matrix with a channel number of $M$, and $I_{M\times 1}$ is the corresponding output power intensities for each sampling channel. Using regression algorithms, the incident spectrum $\Phi$ can be accurately retrieved even when $M$ is considerably smaller than $N$, following:

$$\text{Minimize } \|I - S\Phi\|_2 \text{ subject to } 0 \leq \Phi \leq 1 \tag{4}$$

Eq. (4) is suitable for solving discrete signals (e.g., laser signals). For continuous signals, additional regularization terms can be added to mitigate the ill-conditioning of the undetermined problem, such as the modified Tikhonov regularization[70], modifying Eq. (5) to:

$$\text{Minimize } \|I - S\Phi\|_2 + \alpha\|\Gamma\Phi\|_2 \text{ subject to } 0 \leq \Phi \leq 1 \tag{5}$$

where $\Gamma$ is a difference-operator that calculates the derivative of $\Phi$, and $\alpha$ is the regularization weight that helps suppress noise-induced reconstruction errors. For more complex hybrid incidence (i.e. with both discrete and continuous signals), the Eq. (5) should be further modified by introducing segmented regularization terms, as:

$$\textit{Minimize } \|I - T(\Phi_1 + \Phi_2)\|_2 + \alpha\|\Phi_1\|_1 + \beta\|\Gamma_2\Phi_2\|_2 \text{ subject to } 0 \leq \Phi_1, \Phi_2 \leq 1 \tag{6}$$

where $\Phi_1$ and $\Phi_2$ denote the narrowband and broadband spectral components, respectively; $\alpha$ and $\beta$ are the corresponding weight coefficients. In our study, all spectra reconstruction are performed based on the above equations using the CVX regression algorithm[71].

While our RS demonstrates ultra-high resolution and bandwidth, it is not always suitable to perform every reconstruction using its full capabilities considering not only the computational burden (see Fig. 4f,g), but also the fact that practical incident signals usually only occupy a certain spectral range. Therefore, we adopt a progressive strategy for spectral reconstruction, starting with a coarse reconstruction over the entire bandwidth and moving to a fine reconstruction within a selected spectral window. Specifically, an initial low-resolution, full-band reconstruction is conducted to locate the main energy distribution of the unknown incident spectrum. After identifying its general spectral range, we then narrow the reconstruction window and increase the resolution until the optimal result is achieved. On the other hand, prior knowledge about the incident spectra can be also used to enhance the reconstruction efficiency. For instance, in our NIR sensing experiments, all samples are sequentially illuminated by a group of SLDs, which allows us to adjust the reconstruction window according to the coverage of the corresponding SLD.

**Sample preparation**

The solid samples used in the experiments include ten different types of plastic and coffee. The ten plastic samples cover a range of commonly-used polymer and copolymer materials with different chemical composition and physical properties, such as polycarbonate (PC), polyvinyl alcohol (PVA), acrylonitrile styrene (AS), and others. The ten coffee samples are produced from various regions, such as East Java, Rwanda, and Yunnan, and also vary in tree species and roasting methods. During our testing, all samples are ground and sieved to ensure the uniformity in particle size. The processed particles are then placed into glass containers for reflectance testing.

The liquid samples include the aqueous solutions of glucose and ethanol, and the organic solutions of EG in ISO, all carefully prepared at varying concentrations. The glucose solutions are made by mass percentage, while the ethanol solutions and EG solutions are prepared by volume percentage. The solutions are then transferred into cuvettes, sealed, and tested for absorptance.

**Modelling of NIR spectrometric sensing**

For the reflectance spectra of solid samples, we employ the SVM algorithm to develop classification models. SVM is a supervised learning technique that finds the optimal hyperplane in the feature space to separate data points of different classes. The model performance is evaluated based on classification accuracy, with model's hyperparameters tuned to achieve optimal average results in cross-validation on the training set, thereby minimizing the risk of overfitting. The trained model is then used to classify the plastic samples in the testing set.

Regarding the absorptance spectra of solution samples, RF algorithm is adopted to train prediction models. RF is an ensemble learning method that constructs multiple decision trees to make predictions, where each tree is trained on random subsets of the data to enhance overall predictive performance. Here, we employ the random forest regressor from the Scikit-Learn library in Python for model construction. RMSE is used as the criterion to optimize model's key hyperparameters, such as the number of decision trees and the maximum depth of each tree. By fine-tuning these hyperparameters, we ensure a balance between model complexity and generalization. The optimized model is then applied to predict the concentrations of test samples.

**Data availability**

All data needed to evaluate the conclusions in the paper are present in the paper and/or the Supplementary Materials. A collection of the spectra data in Fig. 6 is available in the University of Cambridge Repository at https://doi.org/10.17863/CAM.110905. Additional inquiries regarding the data can be directed to the corresponding author.

**Code availability**

The spectra reconstruction algorithm, i.e. the CVX convex optimization tool, is available at: https://cvxr.com/cvx/. The code for the SVM and RF models is available from the corresponding author upon request.

**Acknowledgement**

This research was supported by the UK EPSRC through project QUDOS (EP/T028475/1), and also received support from GlitterinTech Limited. The authors thank Mr. Tao Zhang, Mr. Bobo Liu and Ms. Mengting Wu, Mr. Yanlong Liang, and Mr. Yahang Chen for the help in experiments.

**Author contributions**

C.Y. conceived the spectrometer design, performed the optical simulations (with P.B.'s contribution), and drawn the chip layout. W.Zhang performed the characterization of spectrometer performance and analyzed the data with C.Y.'s help. C.Y. designed and conducted the NIR spectroscopic sensing experiments with J.M., W.Zhuo, J.Z., L.M. and T.Y.'s assistance. J.M., M.C., and T.Y. developed the electrical driving broads. C.Y. drafted the manuscript, with W.Zhang, Z. S., and Q.C.'s input. R.P. and Q.C. supervised the project.

**Competing interests**

GlitterinTech Limited declares a pending patent application filed with the China National Intellectual Property Administration (inventors: C.Y. and Q.C., application number:

CN2023107701923), pertaining to the spectrometer design presented in this manuscript. The authors declare no other competing interests.